\documentclass[journal]{IEEEtran}
\usepackage{microtype}
\usepackage{graphicx}
\usepackage{subfigure}
\usepackage{booktabs} 
\usepackage{tabularx}
\usepackage{amsmath,amsfonts,bm}
\usepackage{amsthm, amssymb, bbm}
\usepackage{thmtools}
\usepackage{thm-restate}
\usepackage{lipsum}
\usepackage{mathtools}

\usepackage{hyperref}

\usepackage{graphics} 
\usepackage{graphicx}
\usepackage{array}
\usepackage{epsfig} 
\usepackage{color}
\usepackage{float}
\usepackage[noabbrev]{cleveref}
\usepackage{algorithm}
\usepackage{algorithmic}
\usepackage{url}
\usepackage{calrsfs}
\usepackage{textcomp}
\usepackage{multirow}
\usepackage{booktabs}
\usepackage{textcase}

\usepackage{multirow}
\usepackage{hhline}
\DeclareMathAlphabet{\pazocal}{OMS}{zplm}{m}{n}
\DeclareMathAlphabet{\mathpzc}{OT1}{pzc}{m}{it}
\newcolumntype{P}[1]{>{\centering\arraybackslash}p{#1}}
\newcolumntype{N}{@{}m{0pt}@{}}

\begin{document}

\title{Using Mobility for Electrical Load Forecasting During the COVID-19 Pandemic}
\author{Yize Chen, Weiwei Yang and Baosen Zhang
\thanks{Y. Chen and B. Zhang are with the Department of Electrical and Computer Engineering, University of Washington, Seattle, WA 98195; e-mails: \{yizechen,zhangbao\}@uw.edu. W. Yang is with Microsoft Research, Redmond, WA 98052; e-mail: weiwya@microsoft.com}
}
\maketitle

\begin{abstract}
	The novel coronavirus (COVID-19) pandemic has posed unprecedented challenges for the utilities and grid operators around the world. In this work, we focus on the problem of load forecasting. With strict social distancing restrictions, power consumption profiles around the world have shifted both in magnitude and daily patterns. These changes have caused significant difficulties in short-term load forecasting. Typically algorithms use weather, timing information and previous consumption levels as input variables, yet they cannot capture large and sudden changes in socioeconomic behavior during the pandemic.

	In this paper, we introduce mobility as a measure of economic activities to complement existing building blocks of forecasting algorithms. Mobility data acts as good proxies for the population-level behaviors during the implementation and subsequent easing of social distancing measures. The major challenge with such dataset is that only limited mobility records are associated with the recent pandemic. To overcome this small data problem, we design a transfer learning scheme that enables knowledge transfer between several different geographical regions. This architecture leverages the diversity across these regions and the resulting aggregated model can boost the algorithm performance in each region’s day-ahead forecast. Through simulations for regions in the US and Europe, we show our proposed algorithm can outperform conventional forecasting methods by more than three-folds. In addition, we demonstrate how the proposed model can be used to project how electricity consumption would recover based on different mobility scenarios.

\end{abstract}


\section{Introduction}
The coronavirus disease 2019 (COVID-19) pandemic has impacted almost every aspect of our society and the electric grid is no exception. As many sectors continue to operate remotely through communication technologies, the grid is more than ever operating as the glue that holds the society together at these challenging times. As electric utilities and system operators strive to provide reliable power to communities when they need it the most, the pandemic has caused challenges ranging from the health and safety of frontline crews to the long term supply chains~\cite{Paaso20}. In this paper, we focus on a specific challenge: the unprecedented changes in electricity consumption patterns and the need to provide better load forecasting algorithms.

One of the most striking impacts of the pandemic on the grid is the changes in load consumption patterns and the peak demands. For example, in the United States, as stay-at-home orders were issued by the local and state governments and social distancing were practiced to slow the outbreak of COVID-19, power consumptions shifted both in magnitude and daily patterns. The overall electricity usage has fallen to a 16-year low in the US~\cite{WEF2020} with significant regional variations. Both the PJM and NYISO experienced about 9\% decreases in peak demand in March of 2020~\cite{Paaso20}, with New York City observing decreases up to 21\% in April of 2020~\cite{NYISO20}. Similar shelter-in-place directives were effective in Europe, and Italy saw the biggest reductions of $25\%$ of demand~\cite{IEA20}.

\begin{figure}[ht]
  \centering
  \includegraphics[scale=0.32]{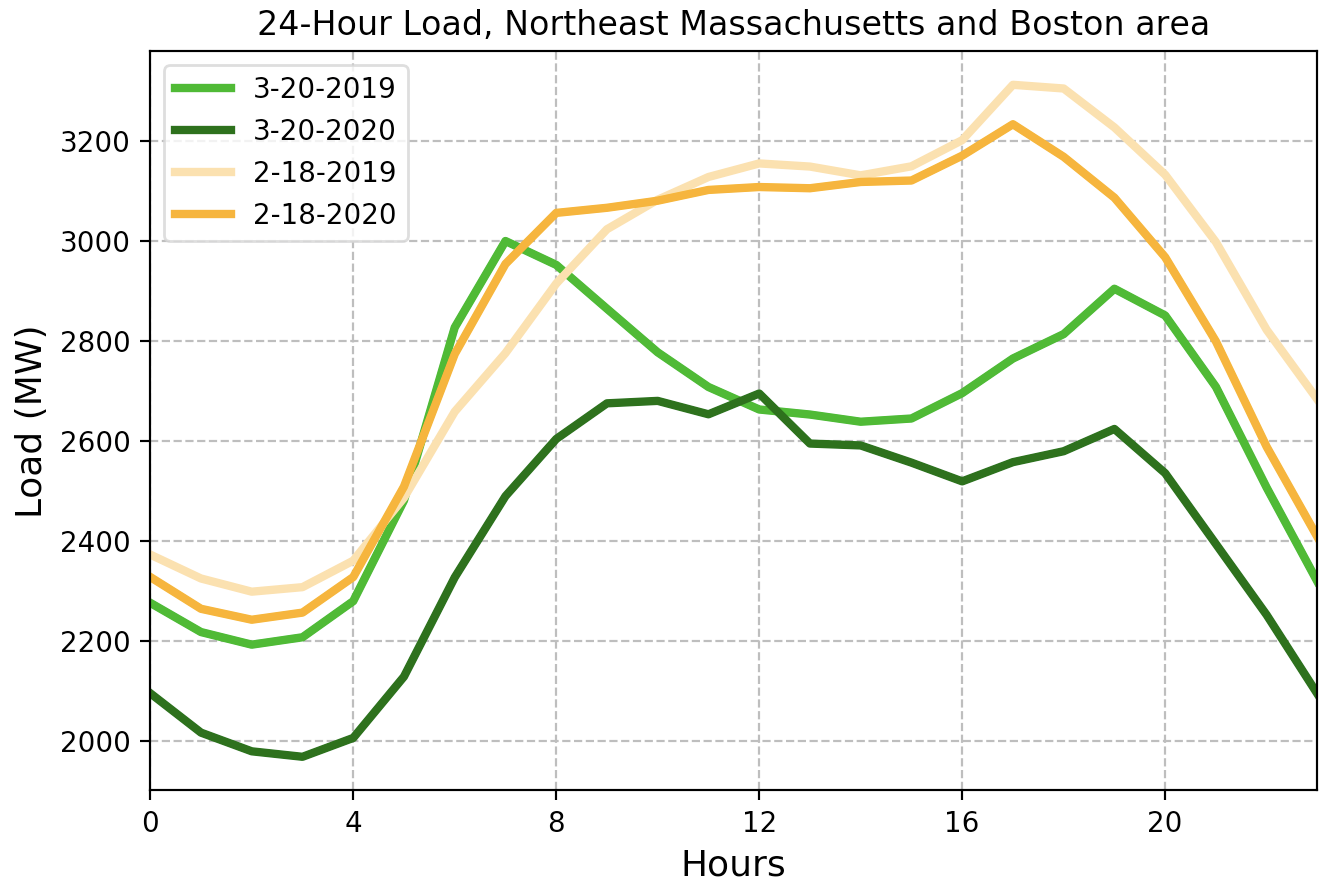}
  \caption{Comparison of changing load patterns of Boston Metropolitan for February 18th and March 20th in 2019 and 2020. All four days are weekdays and experienced similar weather conditions yet with varying load patterns. \vspace{-5pt}}
  \label{fig:NE_load}
\end{figure}

In addition to the changes in peak demand, the temporal patterns in consumption have also shifted to unseen curves as people started to stay at home. For example, Fig.~\ref{fig:NE_load} shows the year-to-year changes for days in February and March for the Boston metropolitan area~\cite{ISONE20} (Boston and the State of Massachusetts stated to enact stay-at-home orders on March 15, 2020). The days are all weekdays with similar weather. The load profiles for Feb 18th, 2019 and Feb 18th, 2020 are relatively similar. However, on March 20, 2019, there is a pronounced double peak pattern, while the same day in 2020 shows a very different pattern with relatively flat load with a small peak in the early evening.

%
%
%
%
%
%

\begin{figure*}
	\centering
	\includegraphics[scale=0.38]{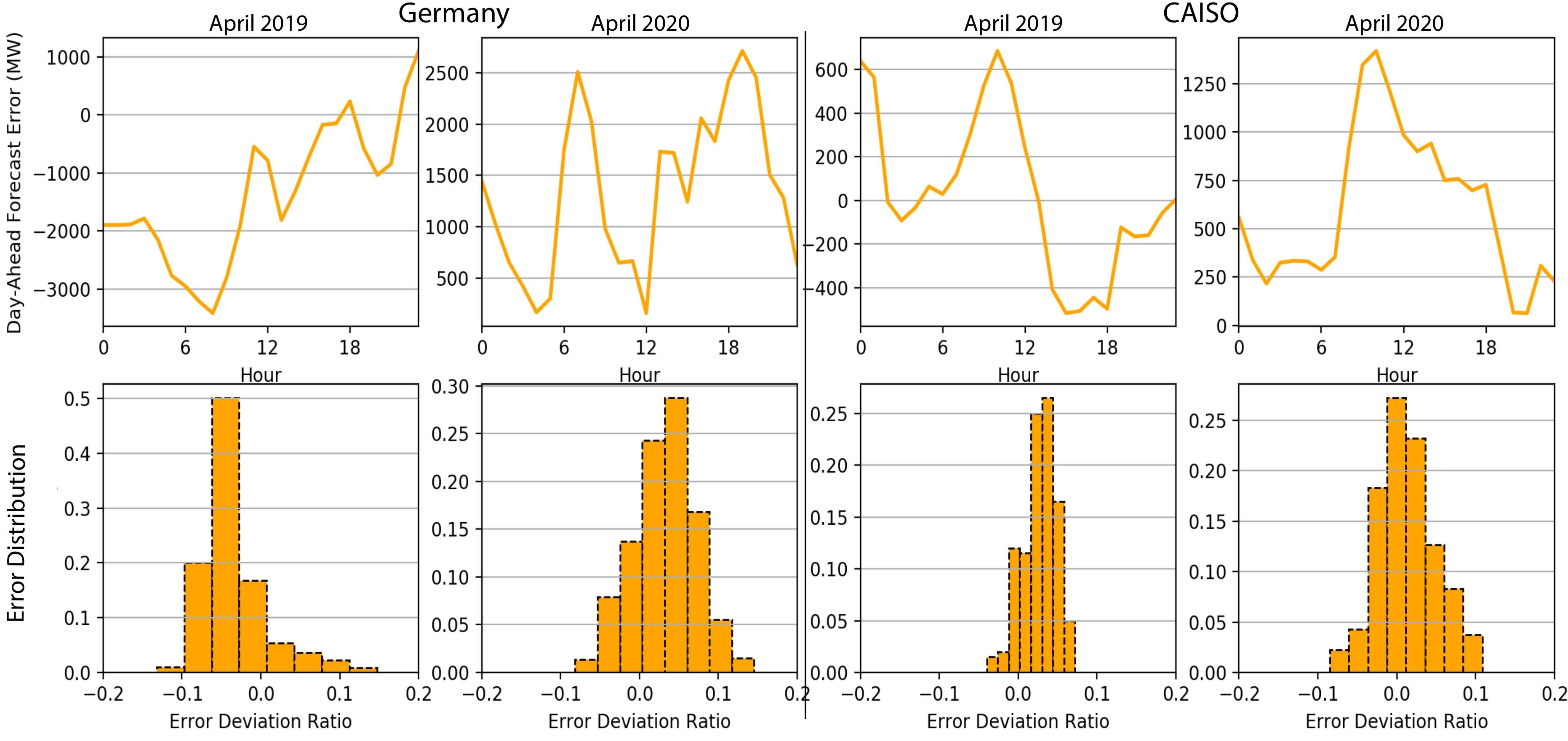}
	\caption{Comparisons of day-ahead load forecasts performance in April 2019 and April 2020 for Germany and California (CAISO), respectively. The upper row shows the magnitudes of the forecasting error for days in April of 2019 and in April of 2020. The bottom row shows the forecast error distribution over April. The errors are significantly larger for April of 2020, even after a month of staying-at-home and adjustment in the load forecasting process. \vspace{-15pt}}
	\label{fig:forecast_error}
\end{figure*}

These sudden and dramatic changes in the consumption pattern has caused difficulties in short-term load forecasting, since no forecasting algorithm could have anticipated these levels of changes in human behaviors. Forecast models are constructed and validated on historical data, while the most important input features for standard load forecasting algorithms are weather, timing information (day of the week, time of the day, seasonal, etc.) and previous consumption data~\cite{gross1987short}. Using these inputs, it is not difficult to construct models---e.g., by training a deep neural network---to achieve a day-ahead forecast error of less than 2\% for a city-scale utility~\cite{park1991electric}. In particular, daily forecasts rely heavily on the assumption that similar days at similar times of the year observe similar load patterns. The COVID-19 crisis fundamentally changes this assumption, as there are no historically similar events ever since the construction of the current grid infrastructure. Therefore it is not surprising that load forecasting algorithms used by operators have encountered much larger forecasting errors since March of 2020.

Forecast errors can be partly mitigated by artificially setting everyday as a weekend day, but this alone is far from sufficient in closing the accuracy gaps~\cite{ISONE20}. Previous research also investigated adaptive learning schemes where different regions or seasons are considered~\cite{fiot2016electricity, grady1991enhancement}, yet no literature have considered the forecast tasks under an unexpected pandemic. Figure~\ref{fig:forecast_error} shows the published day-ahead forecast errors for Germany and CAISO on randomly selected days in April of 2019 and April of 2020.\footnote{Data available from ENTSO-E at https://transparency.entsoe.eu; CAISO at http://www.caiso.com.} There is significant over-forecasting in April 2020 compared to forecasts made in April 2019. Note that at this time Germans and Californians have largely stayed at home for over a month and manual adjustment have been made to the algorithms, but accurate load forecasting is still a challenge for the system operator. In April 2019, Germany reported an under-forecasting error of $3.51\%$ compared to an over-forecasting error of $2.49\%$ in April 2020, while the over-forecasting errors in CAISO increased from $1.28\%$ to $5.39\%$.

Since load forecasting is a fundamental step in power system operations and is used as the basis of decision-making problems such as unit commitment, reserve management, economic dispatch and maintenance scheduling~\cite{gross1987short}. Consequently, the accuracy of forecasted loads directly impact the cost and reliability of system operations~\cite{Hobbs1999}.
As shown in Fig.~\ref{fig:forecast_error}, even if customer behaviors stabilize somewhat, it is hard for forecasting algorithms to adjust and catch up in a short amount of time. This challenge is compounded by the fact that as parts of the world restart their economies in phased approaches, it is likely that customer behaviors would again undergo rapid fluctuations in the foreseeable future~\cite{barker2009macroeconomic}.
Therefore, we need quantitative measure of the social and economic activities to be incorporated in forecasting algorithms, and the measurements should be readily available to help grid operators and utilities with forecasting and understanding of the load patterns over time.

In this paper, we close the gap between on the forecasting performance before and during the pandemic by \emph{introducing mobility data} as a measure of economic activities. The mobility data is used as a complementary component of load forecast model. Population-level mobility data, for instance, transit and shopping trends, show how people are changing their behaviors once distancing mandates are implemented. Such data are readily available from third parties like Apple and Google~\cite{apple,google}, which is location specific and aggregated across the population that enabled mobile services\footnote{There are many other sources of mobility and traffic data, either through a company like INRIX or public transportation departments. These sources, however, contain only data for specific routes or are privately owned.}. Previous research also indicated that mobility is highly correlated with the economic activities in a region~\cite{le2020temporary,Ecola12}, therefore providing a good complementary input for forecasting algorithms.

However, there still exists practical implementation issues once user mobility is taken into account.  We face a small-data problem as most mobility data are only publicly available for parts of 2020. Therefore, we do not necessarily have a lot of counterfactual information, since most of the mobility would be at low levels compared to normal, but we do not know the values of normal data. The key benefit of mobility data is that there is a lot of geographical diversity to offset the lack of temporal data. Different countries in the world, as well as regions in the United States are returning to work in heterogeneous phases~\cite{studdert2020disease,nakamoto2020heterogeneity}. Therefore we propose to obtain enough training samples as well as diversity by combining data from different regions, and to investigate how small changes in mobility would drive changes power consumption. To this end, we design a \emph{transfer learning scheme} that enables knowledge sharing between several regions, and the resulting aggregated model can boost the algorithm performance in each region's day-ahead forecast task~\cite{pan2009survey}.

This study recognizes current difficulties in implementing accurate and reliable load forecast algorithms, and illustrates the need to look at additional features reflecting the electricity usage behaviors during these uncertain times. We demonstrate that an accurate load forecasting, with results as good as those before the pandemic, is achievable by rethinking both the forecasting models' input and architecture. We make the simulation cases along with code publicly available\footnote{https://github.com/chennnnnyize/Load-Forecasting-During-COVID-19}. Specifically, we make the following contributions in this work:
\begin{enumerate}
	\item We identify mobility data as an important complementary component for the forecast task during this global pandemic (Sec.~\ref{sec:setup});
	\item We design a learning algorithm to learn and predict the electricity load by explicitly incorporating the mobility patterns, and further adapt a transfer learning framework to tackle data insufficiency issues (Sec.~\ref{sec:learning});
	\item We perform extensive numerical simulations across various regions and countries to validate forecasting performances, showing they can be greatly enhanced by adopting proposed method. We also provide future projections of electricity demand using our proposed model (Sec.~\ref{sec:simulation}).
\end{enumerate}

\label{sec:intro}

\section{Problem Setup}
\label{sec:setup}
In this section, we first formulate the task of day-ahead load forecasting. We then explain the inherent difficulties of achieving accurate forecasts during periods affected by the COVID-19 pandemic. We advocate that mobility data is a viable candidate feature to satisfy the necessity of incorporating social and economic behaviors in forecasting algorithms.

\subsection{Day-Ahead Load Forecasting}
A host of models have been proposed to fulfill the task of short-term load forecasts which include linear regression, support vector regression, autoregressive models and neural network models (e.g., see~\cite{park1991electric,alfares2002electric,weron2007modeling,hong2016probabilistic} and the references within). Most (deterministic) forecasting algorithms are constructed and fitted to find the mapping from a group of specifically designed input features $\textbf{X}_{t}$ at timestep $t$ to future loads $L_{t+k}$, where $k$ is the forecasting horizon. In this paper, we focus on the task of day-ahead forecasting at hourly resolutions, so $k\in [1,24]$. Without loss of generality, we normalize the loads such that $L_{t+k}\in [0,1]$ for all $k$.

Mathematically, forecasters are interested in finding an accurate model, parameterized by $\theta$:
\begin{equation}
f_\theta(\textbf{X}_{t-H}, ..., \textbf{X}_t)=\hat{L}_{t+k}, \; k=1,\dots,24
\end{equation}
where $H$ determines how much history of training data the operators want to take into consideration for forecasting. Accuracy is normally judged based on the error between the forecasted load and the true load, for example, using the mean squared error (MSE) or mean absolute error (MAE). We assume that a training dataset $\mathcal{D}_{tr}=\{(\textbf{X}_{t-H},..., \textbf{X}_t);L_{t+k}\}_{t=1}^T$ of historical observations is available. The training or model fitting process is to find a $\theta$ that minimizes the error on the training dataset. During implementation, $\theta$ is fixed to predict the future load. We use $f_\theta$ and $f$ interchangeably when the dependence on $\theta$ is clear.


The input features $\textbf{X}_t$ typically include weather, timing information and historical load data~\cite{fiot2016electricity, chen2018short, hong2010modeling}. The weather features are normally considered as the most important drivers of electricity demand, and we include temperature, precipitation, cloud cover and air pressure in this paper. Note that in forecasting, we make use of day-ahead, public-available weather forecasts to construct the input vectors.
 For timing features,  we include one-hot encoded variables (class variables) for hour of the day, month of the year, weekday/weekend distinction and holiday/non-holiday distinction.\footnote{We include the major national holidays with respect to each load region.}
Since there are plenty of historical records of load data and weather measurements and aggregate load data exhibits periodic patterns well explained by the input features, it is not difficult to train an accurate forecasting algorithm for system or regional loads.

\begin{figure}[ht]
	\centering
	\includegraphics[scale=0.4]{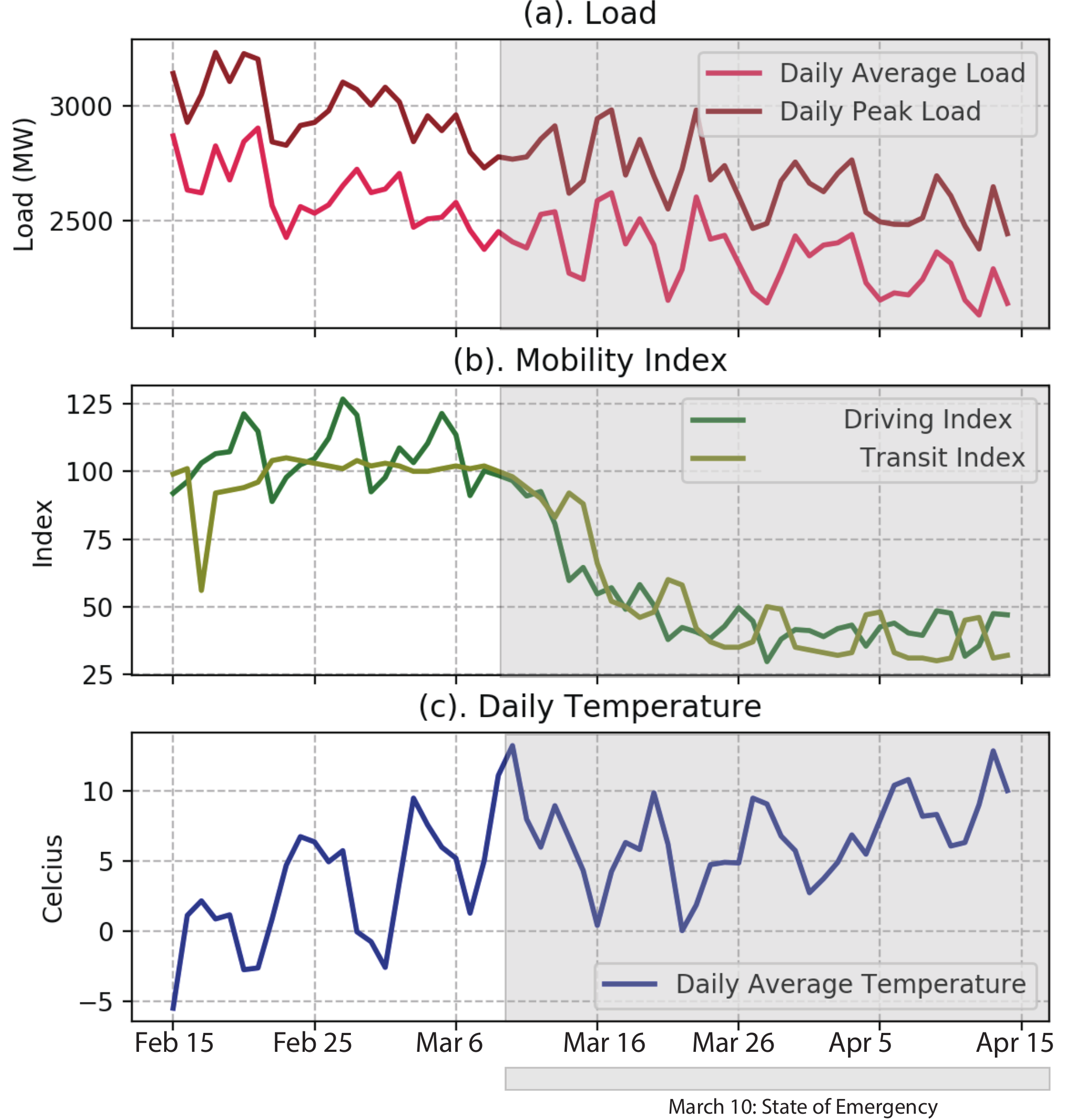}
	\caption{The changing pattern of load, mobility and daily temperatures for Boston area. Daily driving and transit data are from Apple Mobility Trends Reports and Google Community Mobility Reports respectively. \vspace{-10pt} }
	\label{fig:mobility}
\end{figure}

\subsection{Modeling the Effects of COVID-19}
\label{sec:mobility}
Because of economic disruptions and behavioral interventions such as shelter-in-place orders, the forecasting performance using the standard set of features described in the previous section has degraded. This is not unexpected since these features cannot capture the rapid changes in social behaviors and ad hoc fixes such as treating all days like Saturdays have had limited benefit. But inaccurate forecasts could bring significant harm, especially at these times. For instance, overestimation of system loads  may cause generators supplying excessive energy which cause higher grid operation risks, while critical sectors such as hospitals are in need of reliable electricity access.

In this paper, we address two fundamental challenges that arise in forecasting during a pandemic:
\begin{enumerate}
	\item \emph{New features}: Are there other (easily accessible) features that better represent the underlying shifts in social and behavioral patterns?
	\item \emph{Small data}: We only have months of data when the impact of COVID-19 has been evident and social patterns continue to shift somewhat rapidly. How do we design and train a good forecasting algorithm when there is relatively small amount of data?
\end{enumerate}

We find positive answers by introducing a \emph{mobility measure}. Essentially, the forecast algorithm needs to incorporate features that accurately describe how people are changing their behavior in response to COVID-19 pandemic and social distancing policies. For instance, both the spatial and temporal load patterns have changed once people started to work from home.  In this paper, we find a structured and concise representation of \emph{mobility} captures the strong correlation between load and human behavior, and also serves as a valuable feature for subsequent model design. Moreover, with the ubiquitous adoption of smartphones, estimates of mobility are updated frequently and openly available from service providers such as Google and Apple. Other data sources such as stay-at-home population, retail sector population and magnitude of night-time light has been used, but they are more suitable for comparative studies rather than features in a forecasting algorithm~\cite{ruan2020tracking}.

 Fig.~\ref{fig:mobility} shows the load, mobility data and temperature for the Boston metropolitan area from Feb 15th to April 15th, 2020. In Fig.~\ref{fig:mobility} (a), both the peak load and daily average load have decreased over $25\%$, but this pattern is only weakly correlated to increasing temperature (Fig.~\ref{fig:mobility} (c)). In contrast, the load curves exhibits a synchronous decrease with two mobility indexes: the driving index from Apple \cite{apple} and transit index from Google \cite{google}. This is an example of how  mobility measurements can be utilized to take the impacts of pandemic into account to improve load forecasting during COVID-19.

\section{Forecasting Model}
\label{sec:learning}
In this section, we formally describe how we integrate mobility as a socioeconomic feature vector into the forecasting algorithm. We firstly present the architecture of proposed algorithm, followed by a practical implementation during that achieves better performance and generalization by using knowledge transfer between different forecasting tasks.

\subsection{Mobility-Enabled Load Forecasts}
\label{sec:nn_model}
As discussed in Sec. \ref{sec:mobility}, mobility data has the potential to reflect the short-term socioeconomic trends, and we are interested in designing models to flexibly integrate this auxiliary input. We adopt neural networks as the parameterized model to represent $f_\theta$. Neural networks has achieved state-of-the-art performances~\cite{park1991electric, chen2018short, zheng2017electric}, and more importantly, they provide a practical implementation pipeline.
The queries of mobility data can be integrated into neural networks similar to other features such as weather data. Specifically, we concatenate all available features as an input vector, and feed it into the input layer $\textbf{Y}_0=[\textbf{X}_{t-H},\dots, \textbf{X}_t ]$.

While day-ahead weather forecasts are widely available and fairly accurate, there are no mobility forecasts (yet).  Mobility data are often up to real-time observations, so we do a time shift of the corresponding feature spaces to utilize all available inputs at each time step. For instance, to do a day-ahead forecast, we concatenate current day's mobility data along with day-ahead weather forecasts and other class variables as they are all up to date information once the model is implemented.

For the $m$-layer neural network, we parameterize each hidden layer $\mathbf{Y}_{i}, i=1,\dots,m-1$ as a fully-connected layer:
\begin{subequations}
	\label{equ:nn}
	\begin{align}
\mathbf{Y}_{i}=&\sigma_{i}  \left(\mathbf{W}_{i} \mathbf{Y}_{i-1}+\mathbf{b}_{i}\right),\\
\hat{L}_{t+k}=&\mathbf{W}_{m} \mathbf{Y}_{m-1}+b_{m};
	\end{align}
\end{subequations}
where $\mathbf{W}_i,\mathbf{b}_i$ are trainable weights and biases at layer $i$, and $\sigma$ applies elementwise to a vector. The nonlinear activation functions $\sigma_{i}$ promote nonlinearity in the forecasting model and common choices include ReLU and Sigmoid functions~\cite{nair2010rectified}. By collecting the actual load $L_{t+k}$, we use stochastic gradient descent to minimize the mean absolute percentage error (MAPE) during the training process:
\begin{equation}
\label{equ:loss}
L_{MAPE}=\frac{1}{N} \sum_{i=1}^{N} \frac{\left|\hat{L}_{t+k}-L_{t+k}\right|}{L_{t+k}},
\end{equation}
where $N$ is the batch size for model updates. We show in Section~\ref{sec:simulation} that the mobility-augmented input features help to achieve both smaller testing errors and better generalization in comparison to conventional algorithms.

We note the model setup \eqref{equ:nn} and \eqref{equ:loss} are general formulations, which are flexible to the size of input mobility data, the model layers (e.g., recurrent \cite{zheng2017electric} or residual networks~\cite{chen2018short}), and the forecast objective (e.g., point forecasts or probabilistic forecasts~\cite{wang2018combining}).

\begin{figure}[ht]
	\centering
	\includegraphics[scale=0.4]{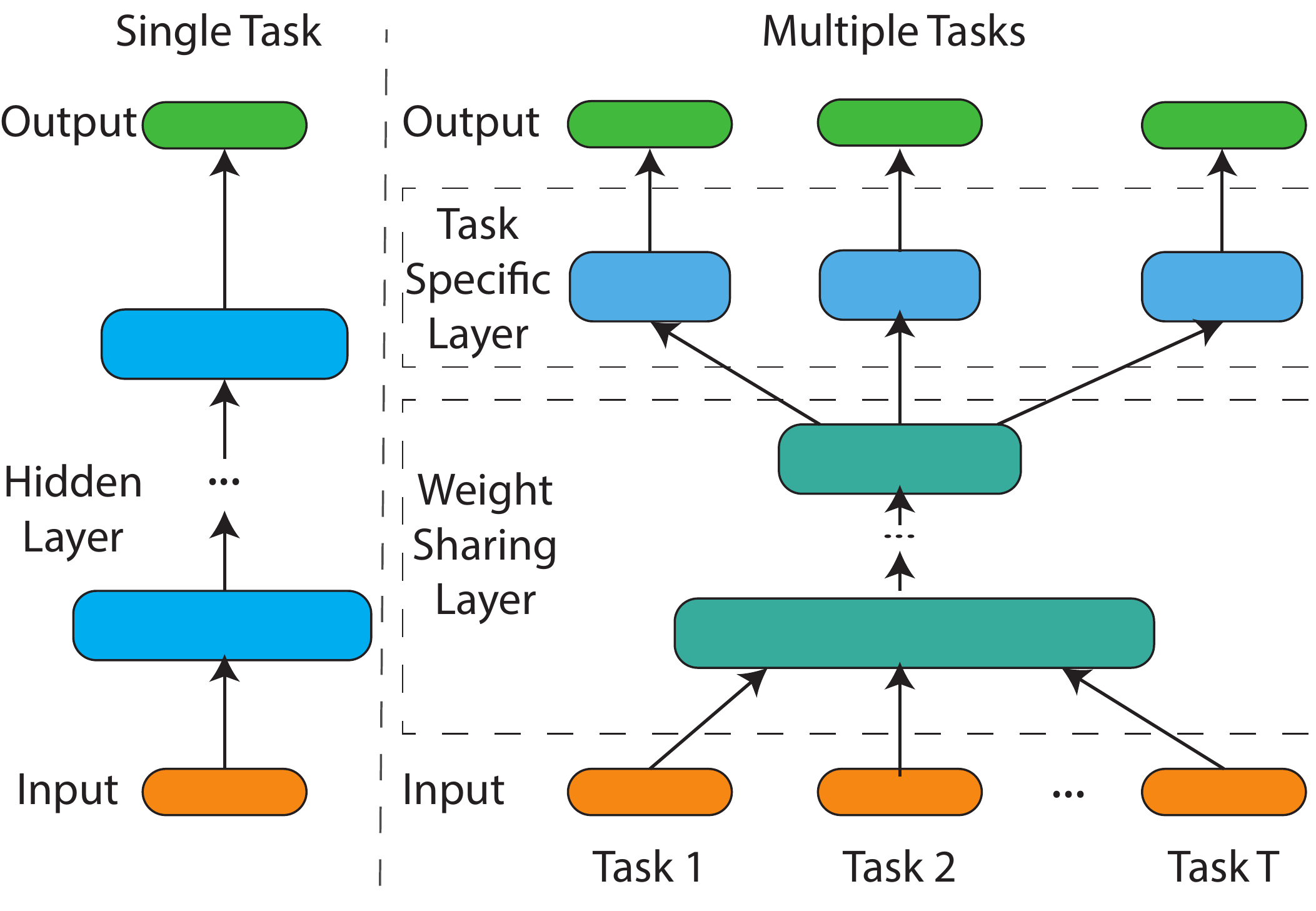}
	\caption{The forecast model architecture for normal forecast task (left) and our proposed model of multiple-task learning (right). \vspace{-20pt}}
	\label{fig:NN}
\end{figure}

\subsection{Knowledge Transfer of Forecasting Tasks}
\label{sec:transfer}
With the incorporation of mobility features, the load forecasting algorithm is expected to have a better handle on the impact of social distancing policies and individual behaviors. Yet there are two challenges in using mobility for load forecasting during COVID-19 outbreaks. The first is that normal forecasting models could use years-long training datasets, while there are only limited electricity consumption and mobility observations since the onset of the pandemic (or the start of social distancing). The second is that mobility and load both dropped sharply in the early stage of stay-at-home orders, but a practical forecasting algorithm should also work when strict social distancing practices are relaxed. It is not obvious whether we would have enough diversity in the data.

We propose to tackle two challenges with one model design, namely, a \emph{multi-task learning framework}~\cite{caruana1997multitask}. This transfer learning procedure is based on the intuition that for related tasks, features useful for one task might be useful for other ones. In our context, consider training forecasting algorithms for two different cities. Normally, these algorithms are trained entirely separately using different datasets. Instead, we explicitly treat these tasks as related to construct the neural network.

We design the multi-task forecasting framework through feature sharing among several different load forecast tasks. Specifically, we co-train several neural networks collectively as illustrated in Fig. \ref{fig:NN}. For a set of forecasting tasks $j=1,\dots, P$ with corresponding collected training datasets $\mathcal{D}_{tr}^{(j)}$, the load forecasting models share the same weights of first $l$ layers, while last $l-m$ layers are mapping the embeddings $\textbf{Y}_l$ to distinct outputs $\hat{L}_{t+k}^{(j)}$. So we are constructing a neural network of the following form:
\begin{subequations}
	\label{equ:transfer}
	\begin{align}
	\mathbf{Y}_{l}=&\sigma_l (\mathbf{W}_{l} (\sigma_{l-1} (\cdots \sigma_{1} (\mathbf{W}_{1} \mathbf{Y}_{0}+\mathbf{b}_1) \cdots))+\mathbf{b}_l),\\
	\hat{L}_{t+k}^{(j)}=&\mathbf{W}_{m}^{(j)} (\sigma_{m-1}^{(j)}  (\cdots \sigma_{l+1}^{(j)}  (\mathbf{W}_{l+1}^{(j)} \mathbf{Y}_{l}+\mathbf{b}_{l+1}^{(j)}) \cdots))+b_{m}^{(j)}.
	\end{align}
\end{subequations}
To train the multi-task forecasting neural network \eqref{equ:transfer}, we iteratively sample a batch of training samples from $\mathcal{D}_{tr}^{(j)}$ from each task $j$ and update the weights for $\mathbf{W}_i, \mathbf{b}_i, \; i=1,\dots, l$ and $\mathbf{W}_i^{(j)}, \mathbf{b}_i^{(j)}, \; i=l+1,\dots, m$. A final fine-tuning step is used to improve the model performance by only training on specific task $j$ while fixing the trained weights $\mathbf{W}_i, \; i=1,\dots, l$.

\begin{figure*}
	\centering
	\includegraphics[width=6in, height=4in]{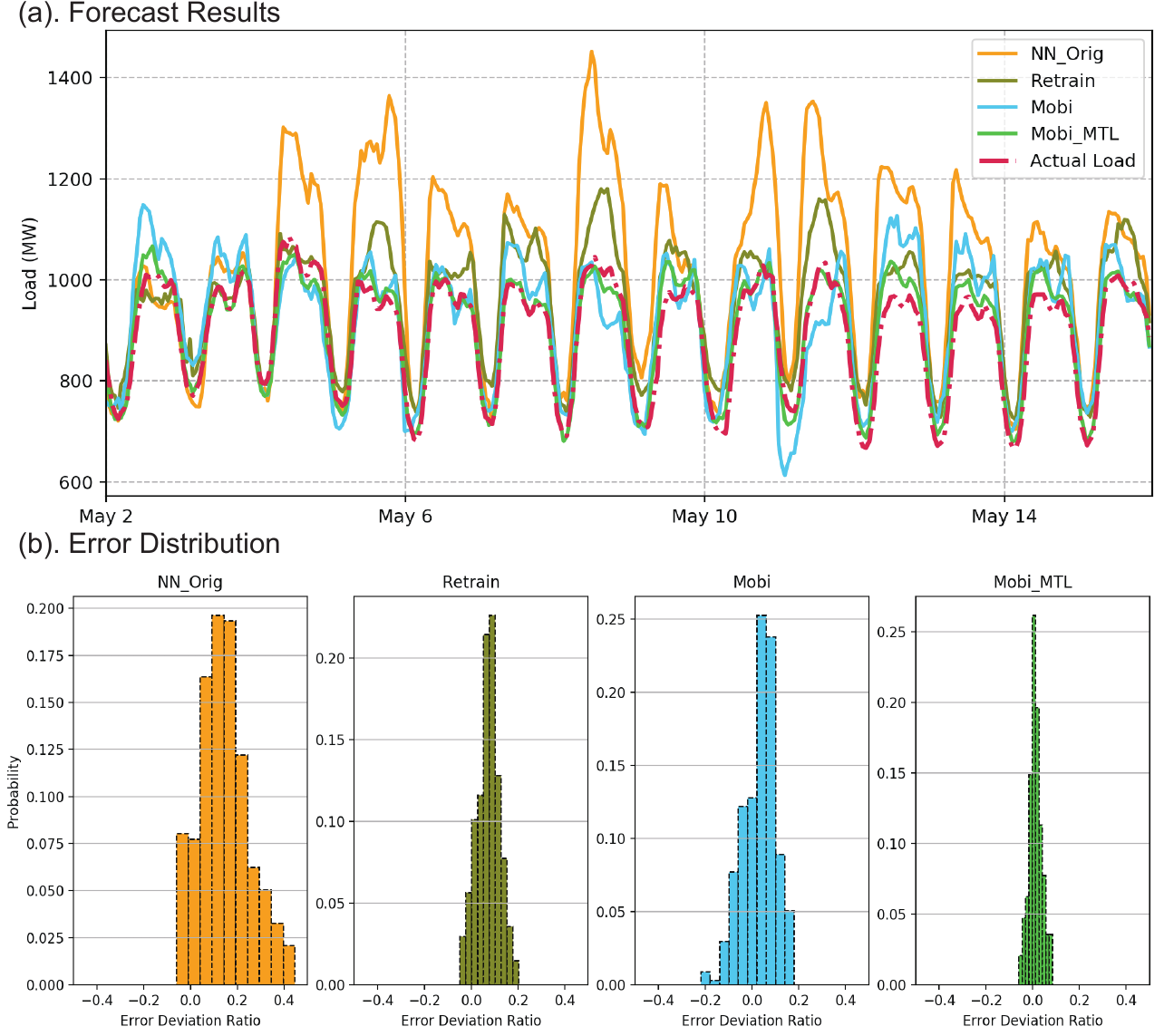}
	\caption{Load forecast results on Seattle City Light dataset. By incorporating mobility data and using multi-task learning procedure, our model achieve lower MAPE error on testing data, while the error distribution is more centered around 0 compared to normally constructed forecast models. The \texttt{Mobi\_MTL} model was collectively trained using data from Seattle, Boston, Chicago and Mid-Atlantic metro areas. \vspace{-10pt}}
	\label{fig:results}
\end{figure*}


The setting of knowledge transfer is especially helpful to load forecasting during the COVID-19 pandemic, as mobility impacts electricity consumption in every region. However, the actual mobility patterns in each region is determined by a complex set of socioeconomic factors and can different widely. This is a \emph{benefit} to the forecasting algorithms, since we can pool data from different regions to create a much \emph{larger and more diverse training set}. The regional differences in how much electricity each end-use sector consumes and the varying effects of COVID-19 mitigation efforts on the sectors are learned and represented by the final task-independent layers. In addition, since different countries and regions (states in the US) are in different stages of lifting the social distancing orders, the multi-task learning framework enables the knowledge transfer so that forecasting results can be accurate even when there is unseen training data for a particular location.
%

The proposed data-driven knowledge transfer scheme enables efficient modeling and learning between different forecasting tasks, as there is no limitations on the number of tasks, neural network architecture or training data size. Note that it is not sufficient to directly aggregate data from each region to train a single forecasting model, since there would be too much averaging and the algorithm would perform poorly for almost all of the individual tasks.  In contrast,  the multi-task learning framework is able to improve each single task's performance, as we show in the next section.

\section{Case Studies}
\label{sec:simulation}
In this section, we conduct extensive simulations on the load forecasting tasks to validate the proposed method can help during the COVID-19 pandemic. In particular, we compare the mobility and transfer learning enabled forecast model with benchmark models. We also provide a planning analysis by considering different social distancing scenarios.

\vspace{-10pt}
\subsection{Dataset and Simulation Platforms}
\subsubsection{Mobility Data} We collect mobility data from Google COVID-19 Community Mobility Reports~\cite{google} and Apple Covid-19 Mobility Trends Reports \cite{apple}. Note that for most of case studies considered in this work, these data starting from middle of January, 2020 include periods both before and during the stay-at-home restrictions due to COVID-19 pandemic.

 \noindent \emph{Google}: The report includes 6 location-specific metrics: retail \& recreation, grocery \& pharmacy, parks, transit stations, workplaces, and residential. The baseline value is the median during the 5-week period from January 3rd to February 6th, 2020. Google calculates these insights based on data from users who have opted-in to Location History for their Google Account.


\emph{Apple}: The mobility metrics include driving, walking and transit, and are a relative volume of directions requests by mobile users per country/region, sub-region or city compared to a baseline volume on January 13th, 2020.

\subsubsection{Electricity Demand Data} We collect and construct load datasets for various regions to evaluate the proposed load forecasting approach. Specifically, we use hourly electricity consumption data for systems of different sizes: country-level data of European countries (United Kingdom, Germany and France), ISO-level data (CAISO, NYISO), zonal data in ERCOT (coastal, north central and south central areas) and metropolitan-level data of US cities (Seattle, Chicago, Boston, the Mid-Atlantic area). The European load data were collected from ENTSO-E, while the US data are publicly available from several ISO and participating utilities. All the collected datasets are available along with our code repo for evaluation. We query weather forecast API World Weather Online~\cite{wwo} and apply data normalization to pre-process each dataset. For larger load regions such as CAISO and European countries, we concatenate several major cities' weather and mobility data as the input feature vectors.

\begin{figure}[h]
	\centering
	\includegraphics[scale=0.36]{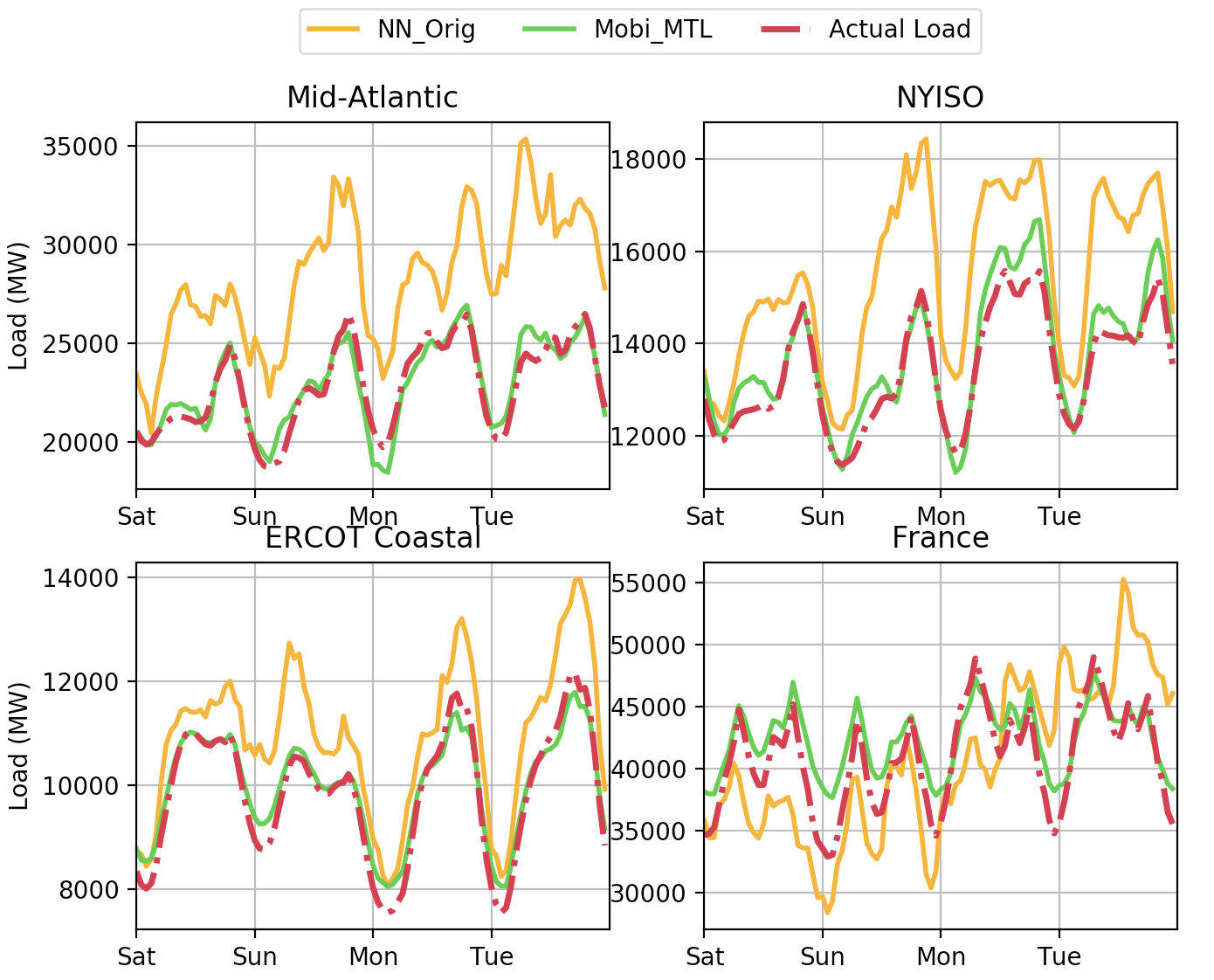}
	\caption{Load forecast performance comparisons on testing samples from different load regions. \vspace{-15pt}}
	\label{fig:load_curve}
\end{figure}

We collect two training datasets to evaluate the proposed approach. The first dataset we consider excludes mobility features, and it covers the time range between January 1st, 2018 to May 15th, 2020. The second dataset makes use of available mobility data ranging from February 15th to May 15th, 2020, which is a relatively small data for load forecasting. This range also spans the pre- and after-lockdown periods for most of the test cases, so it serves as a fair benchmark to evaluate proposed forecasting method.

\subsubsection{Evaluation Methodology}
The following forecasting algorithms are used:
\newline \texttt{NN\_Orig}: This is a standard neural network for day-ahead load forecasting trained on the data from January 1st, 2018 to December 31st, 2019 without mobility data. It acts as a proxy for the actual forecasting model implemented in practice.
\newline \texttt{Retrain}: Using the same model architectures as \texttt{NN\_Orig}, this model simulates the case that model is updated and retrained when COVID-19 pandemic takes impact on electricity loads. We train this model using data ranging from February 15th to April 30th, 2020.
\newline \texttt{Mobi}: Using the techniques described in Sec.\ref{sec:nn_model}, we include the mobility feature vectors, and train this model using data ranging from February 15th to April 30th, 2020.
\newline \texttt{Mobi\_MTL}: This model extends the \texttt{Mobi} model by using the architecture described in Sec.~\ref{sec:transfer}. We train four multi-task learning neural networks based on a selection of similar tasks with similar-sized load regions.

All four models are tested on the electricity load from May 1st to May 15th, 2020, and we report forecast MAPE calculated by \eqref{equ:loss} on the testing dataset. As the underlying regions go through different level of stay-at-home orders during the testing period, we are interested in examining the model's capabilities to predict loads under different socioeconomic scenarios.

\begin{table*}[hbt!]
	\centering
	\begin{small}
		\begin{tabular}
			{P{1.2cm}| P{1.0cm}P{1.0cm}P{1.0cm}P{1.0cm}|P{1.0cm}P{1.0cm}P{1.0cm}|P{1.0cm}P{1.0cm}|P{0.9cm}P{0.9cm}P{0.9cm}}
			\toprule[0.3mm]
			Model&Seattle& Chicago& Boston & Mid-Atlantic &ERCOT Coast&ERCOT NCENT &ERCOT SCENT &NYISO& CAISO&UK & Germany & France \\
			\toprule[0.2mm]
			{\texttt{NN\_Orig}} &15.01 &14.44   &6.55 &14.60 & 7.38& 8.48& 8.16& 12.91& 8.51&10.11&7.73&22.71\\
			{\texttt{Retrain}} &7.55 &17.92   &15.26 &17.27 & 7.17& 9.60& 7.73& 15.55& 7.77&13.78&7.77&8.31\\
			{\texttt{Mobi}} &6.51&4.08&4.38&7.08&1.85&2.70&5.18&6.25&5.97&8.74&6.24&5.93\\
			{\texttt{Mobi\_MTL}} &\textbf{2.28}&\textbf{2.33}&\textbf{2.91}&\textbf{2.61}&\textbf{1.80}&\textbf{1.59}&\textbf{2.71}&\textbf{5.24}&\textbf{3.15}&\textbf{4.46}&\textbf{4.52}&\textbf{4.1}\\

			\bottomrule[0.3mm]
		\end{tabular}
		\caption{Simulation results for day-ahead load forecasts of 12 regions with 4 groups of multi-task learning. \texttt{Mobi\_MTL} is on average 3.98 times better than baseline \texttt{NN\_Orig} model in terms of MAPE across all task benchmarks. \vspace{-10pt}}
		\label{table:results}
	\end{small}
\end{table*}

The simulation platform used in this paper is a laptop with i5-8259U CPU @ 2.30 GHz. The environment used is Python 3.6 with Tensorflow as machine learning package.  To make a fair comparison, we construct load forecasting neural networks with $5$ layers for all tasks along with Dropout regularization, and the first layer has the most number of neurons of $512$. We construct \texttt{Mobi\_MTL} by sharing the weights of first 3 layers across tasks. The dimension of the largest input feature vector in \texttt{Mobi\_MTL} and \texttt{Mobi} for CAISO is 60. For all settings and models, we use $50$ epochs for training. We also note that for the \texttt{Mobi\_MTL} model, even several models are trained together iteratively, the overall training time is less than 20\% of training a single model on the whole dataset which is the case of \texttt{NN\_Orig} model.

\subsection{Evaluation Results}
\subsubsection{Performance on Day-Ahead Load Forecasts}
Forecasted values for the Seattle City Light service region along with the distribution of forecast errors are shown in Fig.~\ref{fig:results} for two weeks' data from May 2nd to May 15th (Seattle's shelter-in-place order were in effect that this time).
 The multi-task learning model is trained using training sets from Boston, Chicago and Mid-Atlantic areas. The \texttt{Mobi} and \texttt{Mobi\_MTL} significantly outperform the other two methods as shown in Table~\ref{table:results}. It can be seen that by integrating mobility data, the trained neural network can better predict the electricity consumption behaviors during this period. The \texttt{Mobi\_MTL} model achieves the smallest forecast error, validating our conjecture that cross-task forecasting knowledge can be helpful when large training datasets are not available.

Table \ref{table:results} compares MAPE on testing datasets are listed for various regions. Each sub-column represents a group of tasks for the multi-task learning setting, while other models are trained separately for each task. The baseline \texttt{NN\_Orig} model results in errors of over $10\%$ MAPE in many cases, much larger than the typical error of 2-4\% before the pandemic. This baseline model is more accurate on weekends than weekdays, indicating that the main changes in load pattern are occurring during workdays. As expected, the original forecast model does not have any knowledge of such drastic socioeconomic changes with the outbreak of COVID-19. The \texttt{Retrain} model can partly redress the overcasting by \texttt{NN\_Orig}, yet since the training dataset for the pandemic period is relatively small, while the mobility patterns are changing frequently during the testing period, such retraining process are still not satisfactory.

Across all forecasting tasks, the inclusion of mobility data can boost the prediction accuracy. The resulting forecast results during COVID-19 pandemic from our \texttt{Mobi\_MTL} model are comparable to ISO's published results before the pandemic. This shows that mobility data are effective proxies for social and economic behaviors.
Regional differences are also notable in the forecasting results. Forecasts on ERCOT regions achieve the best performance, partly due to the fact that Texas has relatively loose restrictions during April and May compared to other US areas under investigation~\cite{Paaso20}. On the contrary, forecasting loads for larger geographical regions (ISO level and country level) are generally harder than the smaller counterparts, as the mobility data and weather data we consider in the simulation do not encompass every possible fine-grained regions.  US metropolitan areas testing group sees the largest model improvement by using multi-task learning model compared to other models, partly because such forecasting tasks only consider smaller load regions, while cities exhibit similar shifts of commercial and residential load once stay-at-home orders were in effect. In Fig. \ref{fig:load_curve}, we visualize the forecast results on various testing cases (one region from each group of multi-task learning) for both weekends and weekdays. It further validates that mobility-enabled forecasting algorithm can capture the intrinsic interplay between human activities and electricity consumption.

It is interesting to observe that both \texttt{Retrain} and \texttt{Mobi} model exhibit overfitting to the small training dataset, as the first testing week's error is much smaller than the second week. This may challenge the practical implementation of such algorithms, as the trained model can not generalize to the future instances. On the other hand, \texttt{Mobi\_MTL} model outputs accurate forecasts throughout the testing period.

We also validate the model performance via the distribution of forecasts deviation compared to actual load. Without considering the recent trends of mobility, \texttt{NN\_Orig} and \texttt{Retrain} model consistently predict greater load values compared to actual loads. On the contrary, the error distributions of \texttt{Mobi} and \texttt{Mobi\_MTL} are more centered around $0$, indicating that mobility features serve as significant inputs.

\subsubsection{Projections on Reopening Scenarios}
\begin{figure}[h]
	\centering
	\includegraphics[scale=0.33]{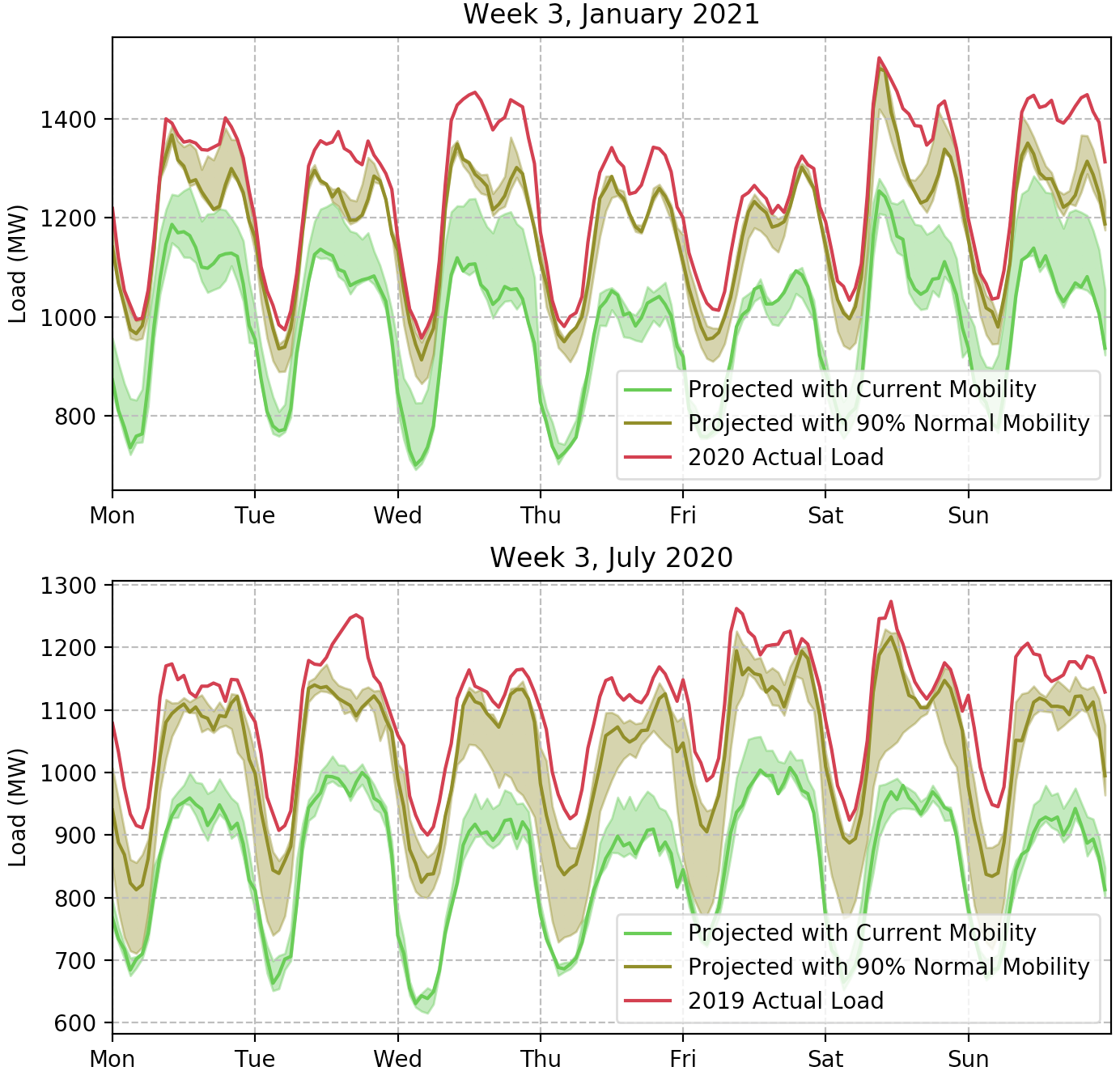}
	\caption{Load projections for January 2021 and July 2020 based on Seattle weather profiles from the previous year considering scenarios of mobility patterns from Week 2, 2020 or from before pandemic dates. Shaded areas plot the load projections based on $95\%$ confidence interval of the mobility data. \vspace{-10pt} }
	\label{fig:future_projections}
\end{figure}

As the objective of research is to provide a fast response to the global pandemic and to prepare the grid for future load changes, we do a future projection analysis based on different mobility patterns by utilizing the proposed load forecasting methodology. The results are illustrated in Fig. \ref{fig:future_projections} for possible load scenarios of Seattle area in July 2020 and January 2021. We plot the two weeks' load curves as representative cases for typical summer and winter load.

According to Washington state's four-phase reopening plan~\cite{WA2020Phase},  the earliest expected date to enter Phase 4 (removal of mandated social distancing orders) is July 13th, 2020. Of course, there are high degrees of uncertainty in both governmental policies and behaviors of people. We analyze two possible future cases based on mobility data. The first case we consider is a recovery of $90\%$ normal mobility. In the second case, we assume there may be extended duration of pandemic impacts, and the mobility pattern stays at the current level. Both the normal mobility and current mobility are estimated using the weekly average values. We concatenate such estimated mobility features along with weather data from the same weeks in July 2019 and January 2020 to construct the input vectors for the \texttt{Mobi\_MTL} model. The $95\%$ confidence interval of mobility data is calculated based on the Gaussian assumption of the estimated mobility indexes, while shaded areas in the figure represent such mobility variations. The model output qualitatively shows the relationship between reduced mobility and reduced load. We can see the decreases of mobility indexes poses more reductions of winter load, with a peak load reduction of over $300MW$ if current mobility patterns persist. We hope by incorporating the projections of mobility, our proposed model can be served as a tool to inform electric grid operators about possible load realizations, and can help system planning in the reopening periods.

\vspace{-12pt}
\section{Discussion and Conclusion}
In this paper, we developed a novel load forecasting method as a timely response to the challenges in load forecasting during the sudden and global COVID-19 pandemic. We discussed approaches to identify the load changes brought by fast-changing socioeconomic behaviors and stay-at-home orders. By explicitly incorporating  mobility patterns, our approach can greatly reduce the error between forecasts and actual loads. We evaluated the proposed approach on load forecasting tasks across a large set of heterogeneous regions globally, and the algorithm not only achieved 3.98 times smaller errors than standard forecasting methods, but also generalized well into varying dates after the outbreak of COVID-19. As the global pandemic may still pose impacts to the power grids in the future, we think techniques developed in the paper could inform grid operators possible future load patterns.


\vspace{-12pt}
\bibliographystyle{IEEEtran}
\bibliography{mybib}

\end{document}